\documentclass{aastex}
\usepackage{emulateapj5,times,mathptm}
\journalinfo{Accepted to AJ April 15, 2004}
\slugcomment{Accepted April 15, 2004}

\shorttitle{
\chandra\ Observations of $z>4$ Radio-Loud Quasars}
\shortauthors{BASSETT ET AL.}

\newcommand{\ltsima}{$\; \buildrel < \over \sim \;$}
\newcommand{\simlt}{\lower.5ex\hbox{\ltsima}}
\newcommand{\gtsima}{$\; \buildrel > \over \sim \;$}
\newcommand{\simgt}{\lower.5ex\hbox{\gtsima}}

\def\lesssim{\mathrel{\hbox{\rlap{\hbox{\lower4pt\hbox{$\sim$}}}\hbox{$<$}}}}
\def\gtrsim{\mathrel{\hbox{\rlap{\hbox{\lower4pt\hbox{$\sim$}}}\hbox{$>$}}}}

\def\arcsec{\hbox{$^{\prime\prime}$}}

\def\aox{$\alpha_{\rm ox}$}
\def\Luv{$L_{\rm UV}$}

\def\Lx{$L_{\rm X}$}

\def\ab1450{$AB_{1450(1+z)}$}

\def\xray{\hbox{X-ray}}

\def\asca{{\it ASCA\/}}
\def\chandra{{\it Chandra\/}}

\def\heao1{{\it HEAO-1\/}}

\def\rosat{{\it ROSAT\/}}

\def\xmm{{\it XMM-Newton\/}}

\begin{document}

\title{Chandra Observations of Radio-Loud Quasars at ${\it z}>4$: \\
X-rays from the Radio Beacons of the Early Universe}

\author{
L.~C. Bassett,\altaffilmark{1,2} W.~N. Brandt,\altaffilmark{1}
D.~P. Schneider,\altaffilmark{1} C. Vignali,\altaffilmark{1,3}
G. Chartas,\altaffilmark{1} and G.~P. Garmire\altaffilmark{1}
}
\altaffiltext{1}{Department of Astronomy and Astrophysics, The
Pennsylvania State University, 525 Davey Laboratory, University
Park, PA 16802, USA; lcb138@astro.psu.edu, niel@astro.psu.edu,
dps@astro.psu.edu, chartas@astro.psu.edu, and garmire@astro.psu.edu.}
\altaffiltext{2}{NASA and NSF-supported undergraduate research associate.}
\altaffiltext{3}{INAF --- Osservatorio Astronomico di Bologna, via Ranzani 1,
40127 Bologna, Italy; vignali@kennet.bo.astro.it}

\begin{abstract}

We present the results of \chandra\ observations of six radio-loud
quasars (RLQs) and one optically bright radio-quiet quasar (RQQ)
at \hbox{\(z\approx\)~4.1--4.4}.  These observations cover a
representative sample of RLQs with moderate radio-loudness
(\hbox{\(R\approx\)~40--400}), filling the \xray\ observational gap
between optically selected radio-quiet quasars (predominantly \hbox{$R
\lesssim 2$--10}) and the five known blazars at $z>4$
(\hbox{$R\approx$~800--27000}), where \hbox{$R$=\(f_{\rm
5~GHz}/f_{\rm 4400~\mbox{\scriptsize\AA}}\)} (rest frame). We
study the relationship between \xray\ luminosity and
radio-loudness for quasars at high redshift and constrain RLQ
\xray\ continuum emission and absorption.  From a joint
spectral fit of nine moderate-$R$ RLQs observed
by \chandra, we find tentative evidence for absorption
above the Galactic $N_{H}$, with a best-fit neutral intrinsic column density of
\hbox{$N_{H}=2.4^{+2.0}_{-1.8}\times 10^{22}$~cm$^{-2}$},
consistent with earlier claims of increased
absorption toward high-redshift RLQs.  We also search for evidence of an
enhanced jet-linked component in the \xray\ emission due to the
increased energy density of the cosmic microwave background (CMB) at
high redshift, but we find neither spatial detections of \xray\
jets nor a significant enhancement in the \xray\ emission relative
to comparable RLQs at low-to-moderate redshifts.  Overall, the
\hbox{$z\approx$~4--5} RLQs have basic \xray\ properties consistent with
comparable RLQs in the local universe, suggesting that the
accretion/jet mechanisms of these objects are similar as well.

\end{abstract}


\section{Introduction}

The study of the $z>4$ universe has become a major theme in
astronomy over the past decade. In particular, \xray\ observations
of quasars at $z > 4$ reveal the physical conditions in the
vicinities of the first massive black holes to form in the
Universe. Prior to 2000, there were only six published \xray\
detections of quasars at $z\ge4$ (see Kaspi, Brandt, \& Schneider 2000;
hereafter KBS00).  In the last four years, through the
unprecedented sensitivities of \chandra\ and \xmm\ as well as optical
wide-area surveys such as the Sloan Digital Sky Survey (SDSS; York
et al.\ 2000), the number has risen to nearly one hundred.

Equally important are wide-area radio surveys, such as Faint
Images of the Radio Sky at Twenty Centimeters (FIRST; e.g., Becker
et al.\ 1995) and the NRAO VLA Sky Survey (NVSS; Condon et al.\
1998), to constrain the properties of radio cores, jets, and
lobes. Radio-selected samples of radio-loud quasars (RLQs),
although sampling a minority of the total quasar population, are
less prone to selection effects than optical surveys, as radio
emission is not affected by absorption due to dust.  The
radio-loudness parameter, as defined by Kellermann et al.\ (1989),
is given by \hbox{$R$=\(f_{\rm 5~GHz}/f_{\rm
4400~\mbox{\scriptsize\AA}}\)} (rest frame), where quasars with
$R\ge10$ are RLQs and those with $R<10$ are radio-quiet quasars
(RQQs).

Some of the first $z>4$ quasars studied in X-rays were a small
group of highly radio-loud blazars\footnote{
The term ``blazar'' is admittedly vague. As defined in \S 1.3 of
Krolik (1999), blazars include BL Lacs and optically violently variable (OVV)
 quasars.  Previous studies have suggested that four of the
$z>4$ quasars discussed in this paper be granted blazar status
(PMN~0525$-$3343, Worsley et al.\ 2004a;
RX~J1028$-$0844, Grupe et al.\ 2004;
GB~1428$+$4217, Worsley et al.\ 2004b;
GB~1508$+$5714, Yuan et al.\ 2003), and we tentatively
place GB~1713$+$2148 (Vignali et al.\ 2003a) in the same category
based on its extreme radio-loudness. }
(\hbox{$R\approx$~800--27000}) in which the \xray\ radiation is
probably dominated by a jet-linked component. These blazars
represent a minority of the RLQ population, however, and so are
not suitable for representative statistical studies. Recently,
{\it Chandra's\/} Advanced CCD Imaging Spectrometer (ACIS) has
obtained \xray\ detections of many RQQs (predominantly with
\hbox{$R \lesssim 2$--10}), which represent the majority
population of quasars in the local universe and so too,
apparently, at high redshift (e.g., Schmidt et al.\ 1995; Stern et
al.\ 2000). Through these studies (e.g., Brandt et al. 2002;
Vignali et al. 2001, 2003a,b), the basic \xray\ properties of
\hbox{$z\approx$~4--5} RQQs have become reasonably well
understood. Perhaps surprisingly, given the strong evolution of
the quasar population (e.g., Boyle et al.\ 2000; Fan et al.\
2001b), high-redshift RQQs have remarkably similar \xray\ and
broad-band properties to RQQs at low-to-moderate redshift (after
controlling for luminosity effects; e.g., Vignali, Brandt, \&
Schneider 2003, hereafter VBS03), suggesting that the accretion
mechanism does not dramatically evolve with time. Prior to the current
paper, however, there were only four published \xray\ detections
of $z>4$ RLQs with \hbox{$R\approx$~10--1000} corresponding to
typical RLQs in the local universe (excluding the blazar
PMN~0525$-$3343 which has $R\approx 800$).

Here we report \chandra\ detections of six
\hbox{$z\approx$~4.1--4.4} radio-selected RLQs and one RQQ (FIRST
0747+2739) which is radio detected and optically bright. These
objects have been selected as prime targets for \xray\
observations based on their bright optical and radio fluxes,
and they have been chosen to represent the majority population of RLQs.  Four
are FIRST-selected objects published in Benn et al.\ (2002), two
are from the Cosmic Lens All Sky Survey (CLASS; Myers et al.\
2003) published in Snellen et al.\ (2001), and one is from the
SDSS, first presented in Anderson et al.\ (2001). Five of the RLQs
have published detections at both 1.4~GHz and 5~GHz and have flat
radio spectra with $\alpha_{\rm r} \approx -0.4$~to $+$0.3
($f_{\nu} \propto \nu^{\alpha}$). In the case of
FIRST~1309$+$5733, the upper limit on the 5~GHz flux density gives
$\alpha_{\rm r} < 0.06$, so it is possible that this object is
also a flat-spectrum RLQ.  None of the quasars shows any
significant structure in the FIRST radio images (these are the
best radio images available, with an angular resolution of
$\approx 5$\arcsec). While the RQQ FIRST~0747$+$2739 is relatively
radio-weak compared to our other objects, it is one of the
optically brightest quasars at $z>4$, and an overabundance of
C~{\sc iv} absorption in the optical spectrum suggests that it may
be a transition Broad Absorption Line (BAL) quasar (Richards et
al.\ 2002).  Given that RLQs are often substantially \xray\
brighter than RQQs with comparable optical luminosity (e.g.,
Zamorani et al.\ 1981; Worrall et al.\ 1987), these objects are
prime targets for \chandra\ snapshot (e.g., \hbox{$\approx$~4--10
ks}) observations. Indeed, all seven of our sources are detected
with a minimum of $\approx$~20 counts.

\figurenum{1}
\centerline{\includegraphics[angle=0,width=8.5cm]{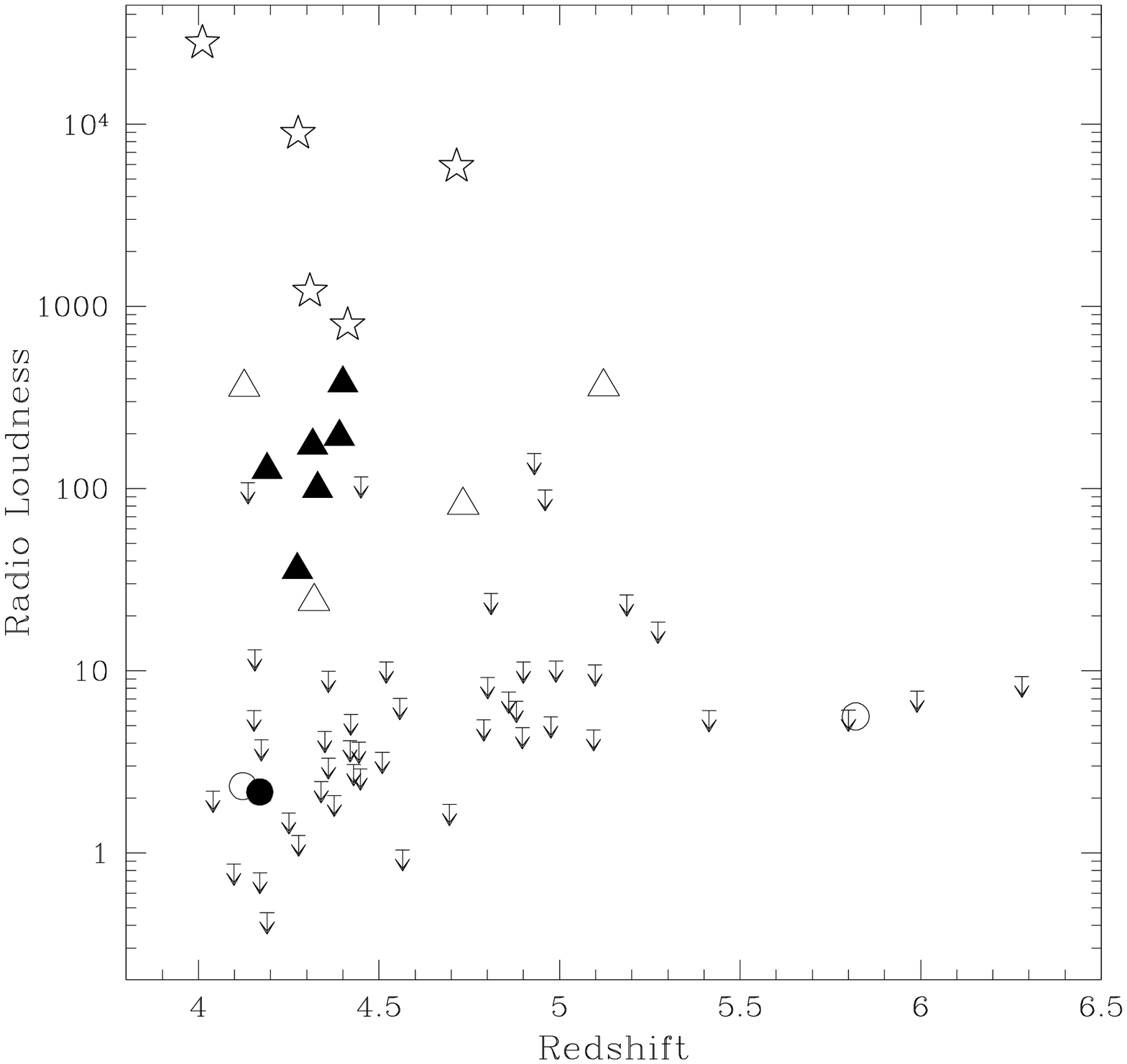}}
\figcaption[bassett.fig1.ps]{\footnotesize Radio loudness versus redshift for
\xray\ detected quasars at $z\geq 4$. The large symbols represent the moderate-$R$
RLQs (triangles), blazars (stars), and RQQs (circles).  The filled symbols
represent the quasars in the current sample.  Arrows show upper limits
for radio loudness at the $\approx 3\sigma$~level based on the
1.4~GHz radio flux density. Note that, excluding
upper limits, the present sample more than doubles the number of
objects with moderate radio-loudness (\hbox{$R \approx$~10--1000}),
filling the \xray\ observational gap between the RQQs and the
blazars. \label{rlvz}} \centerline{} \centerline{}

This brings the total
sample of $z>4$ \xray\ detected RLQs to 15, more than doubling the
number of moderately radio-loud quasars and filling the gap (as
shown in Figure~1) between the many \xray\ observations of $z>4$
RQQs and the extreme blazars.\footnote{
A regularly updated list of $z>4$ quasars with X-ray detections can be
found at
http://www.astro.psu.edu/users/niel/papers/highz-xray-detected.dat}
To date there are $\approx 50$ known $z>4$ RLQs; thus the objects in
our sample comprise $\approx 10$\% of those known.  Our targets have been
drawn from surveys covering a significant fraction of the sky (6400~deg$^{2}$,
Snellen et al.\ 2001; 4030~deg$^{2}$, Benn et al. 2002) and
so are likely to be representative of $z>4$ RLQs as a whole, constituting
the first set large enough to serve as a representative sample for reliable
statistical evaluations of $z>4$ RLQ \xray\ properties.

Recent work (e.g., Cirasuolo et al.\ 2003; Ivezi\'{c} et al.\
2002, 2004) has sparked new controversy over the existence of
bimodality in the radio-loudness distribution of quasars,
suggested first by Strittmatter et al.\ (1980).  Current results
suggest that, although quasars cover the full range of
radio-loudness, the underlying distribution does consist of two
regimes with a local minimum in between.  As reported by
Ivezi\'{c} et al.\ (2002) for a sample of 3225 SDSS quasars
matched to FIRST sources, $8\% \pm 1\%$ of all quasars with
\hbox{$i^{\ast}<$~18.5} are radio-loud, apparently independent of
optical luminosity or redshift, in the range $z<2$. Figure~2 shows
the radio-loudness distribution for $z>4$ quasars with radio detections
or upper limits in FIRST, with the best-fit distribution from Ivezi\'{c} et al.
(2004) overlaid (dashed curve). While the blazars lie on the
extreme radio-loud tail of the distribution, the RLQs from this
work fall near the peak of the radio-loud regime, likely
representing the majority population of $z>4$ RLQs.  By integrating
the best-fit curve directly, we find that our sample of RLQs with
$R=40$--400 is representative of $\simgt 50$\% of the total RLQ population,
while the blazars with $R>1000$ represent $\simlt 5$\% of all RLQs.

\figurenum{2}
\centerline{\includegraphics[angle=0,width=8.5cm]{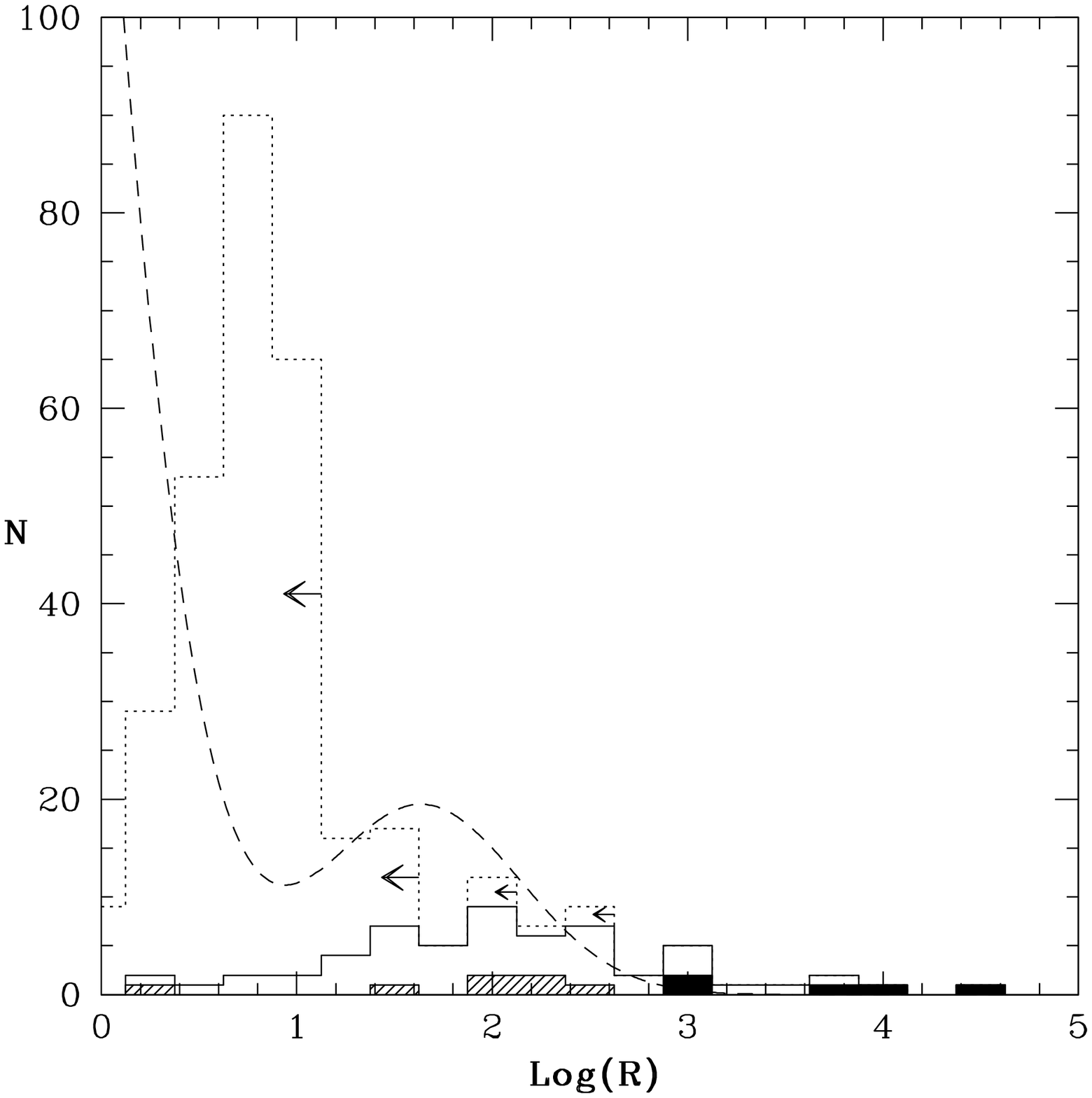}}
\figcaption[bassett.fig2.ps]{Radio-loudness distribution for
known $z \geq 4$ quasars with detections or upper limits
in the FIRST survey.  The solid histogram shows all $z \geq
4$ quasars with radio detections, and the dotted histogram shows
all those with radio upper limits.  The dashed curve is adapted
from the radio-loudness distribution in Ivezi\'{c} et al. (2004),
determined from a sample of $\sim 10,000$ objects detected by
FIRST and SDSS (with arbitrary normalization). The objects in the
current sample (hatched shading) represent the majority population
of RLQs, while the blazars (solid shading) lie on the extreme
radio-loud tail. \label{rlhist}} \centerline{} \centerline{}

Throughout this paper we adopt $H_{0}$=70 km s$^{-1}$ Mpc$^{-1}$
in a $\Lambda$-cosmology with $\Omega_{\rm M}$=0.3 and
$\Omega_{\Lambda}$=0.7 (e.g., Spergel et al.\ 2003).

\begin{table*}
\footnotesize
\tabletypesize{\footnotesize}
\caption{\chandra\ Observation Log}
\begin{center}
\begin{tabular}{lccccccc}
\tableline
\tableline
\multicolumn{1}{c}{Object} &  & FIRST & FIRST & $\Delta_{\rm Rad-X}$\tablenotemark{a} &
\xray & Exp.~Time$^{\ \rm b}$ & \\
\multicolumn{1}{c}{Name} & $z$ & $\alpha_{2000}$ & $\delta_{2000}$ & (arcsec) &
Obs.~Date & (ks) & Ref. \\
\tableline
FIRST~0725$+$3705 & 4.33 & 07 25 18.26 & $+$37 05 18.3 & 0.2 & 2002 Dec 17 & 4.89 & (1) \\
FIRST~0747$+$2739 & 4.11 & 07 47 11.20 & $+$27 39 04.0 & 0.9 & 2002 Dec 03 & 4.96 & (1,3) \\
SDSS~0839$+$5112 & 4.39 & 08 39 46.20 & $+$51 12 02.9 & 0.3 & 2003 Jan 23 & 4.90 & (2,4) \\
CLASS~J0918$+$0637 & 4.19 & 09 18 24.39 & $+$06 36 53.3 & 0.2 & 2002 Dec 10 & 4.90 & (2) \\
FIRST~1309$+$5733 & 4.27 & 13 09 40.69 & $+$57 33 10.0 & 0.4 & 2003 Jul 07 & 4.64 & (1) \\
CLASS~J1325$+$1123 & 4.40 & 13 25 12.48 & $+$11 23 30.0 & 0.5 & 2003 Mar 02 & 4.70 & (2) \\
FIRST~1423$+$2241 & 4.32 & 14 23 08.24 & $+$22 41 58.2 & 0.2 & 2002 Dec 02 & 4.70  & (1) \\
\tableline
\end{tabular}
\vskip 2pt
\parbox{5.3in}
{\small\baselineskip 9pt \footnotesize \indent {\sc Note. ---}
Radio and optical identifications
of the quasars presented here
can be found in the papers cited in the reference column.  Radio
positions are from the FIRST catalog (Becker et al.\ 1995). Units of right ascension are
hours, minutes, and seconds, and units of declination are degrees, arcminutes, and arcseconds. \\
$^{\rm a}$ Distance between the radio and \xray\ positions.
$^{\rm b}$ The \chandra\ exposure time has been corrected for detector dead time. \\
{\sc References. ---} (1) Benn et al.\ 2002; (2) Snellen et al.\
2001; (3) Richards et al.\ 2002; (4) Anderson et al.\ 2001. } 
\end{center}
\vglue-0.9cm
\end{table*}
\normalsize

\section{Observations and Data Reduction}

\subsection{Chandra Observations and Basic Data Reduction}

The \xray\ observations were acquired by \chandra\ during Cycle 4,
using the Advanced CCD Imaging Spectrometer (ACIS; Garmire et al.
2003) with the S3 CCD at the aimpoint.  The observation log is
reported in Table 1.  Faint mode was used for the event telemetry
format, and the \asca\ grade 0, 2, 3, 4, and 6 events were used in
the analysis.  With one exception, the observations are free from
background flares.  A flare occurred near the end of the
observation of SDSS~0839$+$5112, resulting in  an increased
background count rate (by a factor of $\approx 10$) for $\approx
10$~minutes.  Due to the superb imaging capabilities of \chandra,
this flare does not have a substantial effect on our data; the
count rates used in deriving physical quantities have been
background subtracted, and, especially since SDSS~0839$+$5112 is
the \xray\ brightest source of our sample, the additional
background does not significantly affect our spectral analysis.

Source detection was carried out with {\sc wavdetect} (Freeman et
al. 2002).  For each image, we calculated wavelet transforms
(using a Mexican hat kernel) with wavelet scale sizes 1, 1.4, 2,
2.8, and 4 pixels.  Those peaks whose probability of being false
were less than the threshold of \hbox{10$^{-5}$} were declared
real.  This threshold is entirely appropriate when, as is the case
here, the source position of interest is specified {\it a priori}.
The \chandra\ positions of the detected quasars lie
within 0.2--0.9\arcsec of their FIRST radio positions (see Table 1),
consistent with the expected positional error.  Source detection
was performed in four energy ranges:  the ultrasoft band (\hbox{0.3--0.5~keV})
, the soft band (\hbox{0.5--2~keV}), the hard band (\hbox{2--8~keV}) and the
full band (\hbox{0.5--8~keV}).  In the redshift range \hbox{$z\approx$~4.1--4.4}
for these objects, the full band corresponds to the
\hbox{$\approx$~2.5--43}~keV rest-frame band.

\begin{table*}
\footnotesize
\begin{center}
\caption{X-ray Counts, Hardness Ratios, Band Ratios, and Effective Photon Indices}
\tabletypesize{\footnotesize}
\begin{tabular}{lccccccc}
\tableline
\tableline
  & \multicolumn{4}{c}{X-ray Counts\tablenotemark{a}} \\
\cline{2-5} \\
Object & [0.3--0.5~keV] & [0.5--2~keV] & [2--8~keV] & [0.5--8~keV] & Hardness Ratio\tablenotemark{b} & Band Ratio\tablenotemark{b} & $\Gamma$\tablenotemark{c} \\
\tableline
FIRST~0725$+$3705 & $<3.0$ & 21.2$^{+5.7}_{-4.6}$ & {\phn}4.9$^{+3.4}_{-2.1}$ & 26.1$^{+6.2}_{-5.1}$ & $-$0.62$^{+0.19}_{-0.14}$ & 0.23$^{+0.16}_{-0.11}$ & 2.1$^{+0.6}_{-0.5}$ \\
FIRST~0747$+$2739 & $<4.8$ & 16.8$^{+5.2}_{-4.0}$ & {\phn}5.0$^{+3.4}_{-2.1}$ & 21.7$^{+5.7}_{-4.6}$ & $-$0.54$^{+0.22}_{-0.16}$ & 0.30$^{+0.21}_{-0.15}$ & 1.8$^{+0.6}_{-0.5}$ \\
FIRST~1309$+$5733 & $<4.7$ & 24.8$^{+6.1}_{-4.9}$ & {\phn}6.9$^{+3.8}_{-2.6}$ & 31.7$^{+6.7}_{-5.6}$ & $-$0.57$^{+0.18}_{-0.14}$ & 0.28$^{+0.16}_{-0.12}$ & 1.8$^{+0.5}_{-0.4}$ \\
FIRST~1423$+$2241 & $<3.0$ & 16.8$^{+5.2}_{-4.1}$ & {\phn}2.9$^{+2.9}_{-1.6}$ & 19.7$^{+5.5}_{-4.4}$ & $-$0.79$^{+0.22}_{-0.12}$ & 0.12$^{+0.16}_{-0.08}$ & 2.6$^{+1.0}_{-0.8}$ \\
CLASS~J0918$+$0636 & {\phn}5.0$^{+3.4}_{-2.2}$ & 41.6$^{+7.5}_{-6.4}$ & 13.7$^{+4.8}_{-3.7}$ & 55.4$^{+8.5}_{-7.4}$ & $-$0.50$^{+0.13}_{-0.11}$ & 0.33$^{+0.12}_{-0.10}$ & 1.8$\pm 0.3$ \\
CLASS~J1325$+$1123 & $<4.5$ & 23.7$^{+5.9}_{-4.8}$ & {\phn}6.8$^{+3.8}_{-2.5}$ & 30.6$^{+6.6}_{-5.5}$ & $-$0.55$^{+0.18}_{-0.14}$ & 0.29$^{+0.17}_{-0.12}$ & 1.8$^{+0.5}_{-0.4}$ \\
SDSS~0839$+$5112 & {\phn}2.0$^{+2.6}_{-1.3}$ & 55.0$^{+8.5}_{-7.4}$ & 20.6$^{+5.6}_{-4.5}$ & 75.4$^{+9.7}_{-8.7}$ & $-$0.45$^{+0.11}_{-0.10}$ & 0.38$^{+0.11}_{-0.10}$ & 1.6$^{+0.3}_{-0.2}$ \\
\tableline
\end{tabular}
\vskip2pt
\parbox{6.0in}
{\small\baselineskip 9pt \footnotesize \indent
$^{\rm a}$ Errors on the \xray\ counts were computed according to Tables~1 and 2 of
Gehrels (1986) and correspond to the 1$\sigma$~level; these were
calculated using Poisson statistics. The upper limits are at the
95\% confidence level and were computed according to Kraft,
Burrows, \& Nousek (1991).

$^{\rm b}$ Errors on the hardness ratios [defined as
$(H-S)/(H+S)$, where $S$ is the soft-band (\hbox{0.5--2~keV}) counts and
$H$ is the hard-band (\hbox{2--8~keV}) counts], the band ratios ($H/S$),
and the effective photon indices are at the
$\approx$~$1\sigma$~level and have been computed following the
``numerical method'' described in $\S$~1.7.3 of Lyons (1991). This
avoids the failure of the standard approximate variance formula
when the number of counts is small (see $\S$~2.4.5 of Eadie et al.
1971).

$^{\rm c}$ The effective photon indices (and errors) have been
computed from the band ratios and their respective errors, using
the {\sc PIMMS} software.  We have applied a time-dependent
correction which accounts for the quantum-efficiency degradation
of \chandra\ ACIS at low energies (see $\S$3). }
\end{center}
\vglue-0.9cm
\end{table*}
\normalsize

The {\sc wavdetect} \xray\ photometry is shown in Table~2.  We
have checked these results against manual aperture photometry and
found good agreement between the two methods.  Table~2 also lists
the hardness ratios [defined as \((H - S)/(H + S)\), where $S$ is
the soft-band (\hbox{0.5--2~keV}) counts and $H$ is the hard-band (\hbox{2--8
keV}) counts], the band ratios $(H/S)$, and the effective \xray\
power-law photon indices ($\Gamma$) derived from these band ratios
assuming Galactic absorption.  This derivation has been performed
using the \chandra\ \xray\ Center Portable, Interactive,
Multi-Mission Simulator ({\sc pimms}; Mukai 2002) software, and we
have applied a time-dependent correction for the
quantum-efficiency degradation of \chandra\ ACIS at low energies
(see $\S$3 for details). All of the quasars have effective photon
indices consistent, within the significant errors, with previous
studies of RLQs at \hbox{$z\approx$~0--2} (\hbox{$\Gamma \approx$~1.4--1.9};
e.g., Reeves \& Turner 2000).  We also performed basic spectral
analyses for all of the sources using the Cash statistic (Cash
1979) with {\sc xspec} (Version~11.2.0; Arnaud et al.\ 1996) in the
\hbox{0.5--8~keV} band. We found general agreement between the two
methods.  Further spectral analysis is presented in $\S$3.2.

Figure~3 shows the soft-band images (chosen to give the best signal-to-noise
ratio) of the six RLQs, both as raw 0.492$\arcsec$ pixels and
adaptively smoothed at the $3\sigma$\ level using the algorithm of
Ebeling, White, \&~Rangarajan (2004).

\begin{figure*}[!t]
\figurenum{3}
\centerline{\includegraphics[angle=0,width=\textwidth]{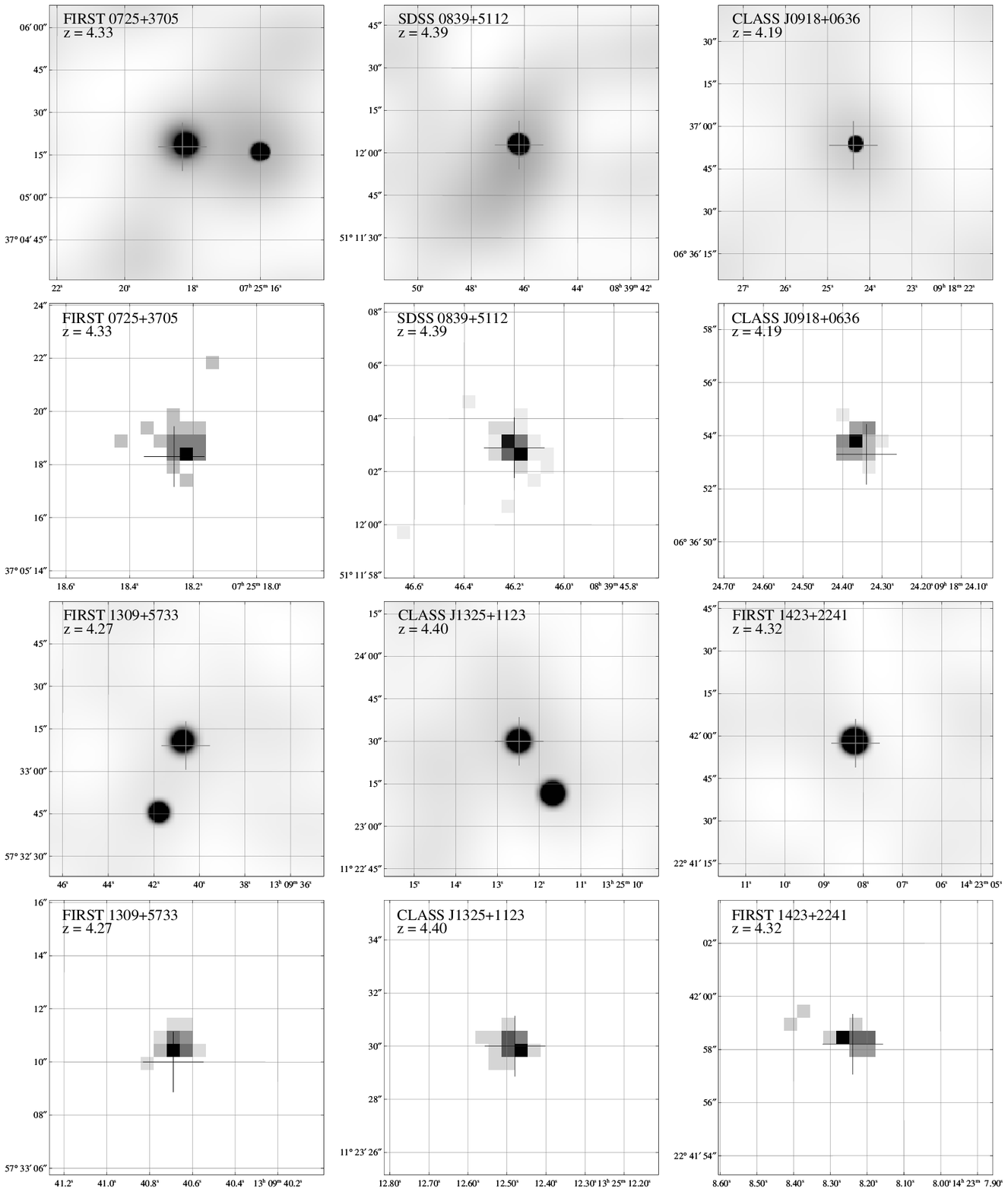}}
\vskip-4.0cm
\figcaption[bassett.fig3.ps]{\chandra\ \hbox{0.5--2~keV} (\hbox{$\approx2.5$--10~keV}
rest frame) images of the
six RLQs presented in this paper. In each panel,
the horizontal axis shows the Right Ascension, and the vertical
axis shows the Declination (both in J2000 coordinates). North is
up, and East to the left.
For each pair, the upper images are 97\arcsec $\times$ 97\arcsec\ and have been
adaptively smoothed at the 3$\sigma$~level [except for
FIRST~1309$+$5733 (2$\sigma$~level)] using the algorithm of
Ebeling et al. (2004).  The lower images are 10\arcsec $\times$ 10\arcsec,
showing raw 0.492\arcsec\ pixels.  Crosses mark the optical positions of the
quasars. \label{images}}
\end{figure*}

\subsection{Companion X-ray Sources}

Powerful \xray\ jets have been observed emanating from many
RLQs,\footnote{
A list of X-ray detected quasar jets is available at
http://hea-www.harvard.edu/XJET/}
 and their presence has been inferred in high-redshift RLQs
(particularly the blazars, in which the jet is presumably oriented
close to the line of sight) from high luminosity and variability
in both the \xray\ and radio bands (e.g., Fabian et al.\ 1997, 1999).
At high redshift such investigations are particularly interesting,
since if a significant component of the \xray\ radiation is a
consequence of inverse Compton (IC) scattering of cosmic
microwave background (CMB) photons, then the \xray\ emission may
be significantly enhanced (see $\S$3.3 for details). Additionally,
in the favored cosmological models, the angular size increases
with redshift beyond \hbox{$z \approx 1.6$} such that the sub-arcsecond
spatial resolution of \chandra\ is sufficient to resolve kpc-scale
jets at all redshifts. Recently, \xray\ jets have been discovered
associated with the \hbox{$z=3.8$} radio galaxy 4C41.17 (Scharf et al.\
2003) and with the \hbox{$z=4.3$} blazar GB~1508$+$5714 (Siemiginowska et
al.\ 2003; Yuan et al.\ 2003).  Schwartz (2004) also presents convincing
evidence for the detection of an IC/CMB enhanced jet associated with
the $z=4.01$ blazar GB~1713$+$2148 in a \rosat\ {\sc HRI} observation.

We have searched for possible companions and jets over a region of
\hbox{$\approx 100\arcsec \times 100\arcsec$} centered on each quasar
(the same regions shown in the smoothed images in Fig.~\ref{images}).  At \hbox{\(z
\approx\)~4.1--4.4}, 100\arcsec\  corresponds to a linear scale of
$\approx 700$~kpc.  In the seven fields,
we found a total of three objects in the soft band,
which can be seen in Fig.~\ref{images}.  None of these is sufficiently \xray\
bright and close to our targets to contaminate their \xray\
emission. The three serendipitous detections visible in
Fig.~\ref{images} all have \hbox{$B$-band} counterparts in the second Palomar
Optical Sky Survey (POSS~II) plates,\footnote{ Available at
http://www.nofs.navy.mil/data/fchpix/ }
 which probably precludes them from being companion objects to the quasars at
high redshift.  In addition, we detected a faint (3 counts) \xray\
source in the hard (\hbox{2--8~keV}) band $\approx$~34\arcsec\ to the
southeast of SDSS~0839$+$5112, which appears to have faint
counterparts in both the POSS~II $R$-band and FIRST images (no
$B$-band counterpart). However, since no counts were detected at
its position in the soft band and jets typically have fairly soft
\xray\ spectra (\hbox{$\Gamma \approx 1.6$--2.3}; Harris \&
Krawczynski 2002), it appears unlikely that the
source is a jet from SDSS~0839$+$5112.

As a further observational check, we compared the \hbox{0.5--2~keV} cumulative
number counts of the sources detected in the quasar fields (the
targeted objects were, of course, excluded) with those presented
by Moretti et al.\ (2003) using a compilation of six different
surveys.  The surface density of \xray\ sources at our observed
flux limit (\hbox{$\approx 2\times 10^{-15}$}~erg cm$^{-2}$ s$^{-1}$) is
\hbox{$N(>S) = 590^{+578}_{-320}$~deg$^{-2}$}, consistent within the
errors with the results (490~deg$^{-2}$ at this flux limit) of
 Moretti et al.\ (2003).  Thus we find no evidence for
an excess of companion objects associated with our quasars.
\subsection{X-ray Extension of the Quasars}

To constrain the presence of either gravitational lensing (e.g.,
Wyithe \& Loeb 2002; Comerford, Haiman, \& Schaye 2002) or \xray\
jets close to the sources, we performed an
analysis of \xray\ extension for all the quasars in the present
sample.  For each quasar, we created a point spread function (PSF)
at \hbox{$\approx$~1.5} keV at the source position (using the {\sc
CIAO} tool {\sc MKPSF}) normalized by the observed number of source counts.
Then we compared the radial profile of this PSF with that of the
source, finding agreement for each object.  To identify a
putative jet by this method,
we require a minimum of $\approx 3$ nearby counts offset
by $\approx 2\arcsec$ from the core, corresponding to a distance
of $\approx 15$~kpc at $z=4.1$--4.4.  We can
exclude significant X-ray extension above this scale for the
objects in the present sample.

\subsection{X-ray Variability}

Given tentative reports of increasing quasar \xray\ variability
with redshift (in the sense that quasars of the same \xray\
luminosity are more variable at \hbox{$z>2$}; Manners, Almaini, \&
Lawrence 2002), we have searched for \xray\ variability by
analyzing the photon arrival times in the \hbox{0.5--8~keV} band
using the Kolmogorov-Smirnov (KS) test. While no significant variability
was detected (the KS probabilities indicate that a constant
hypothesis can only be rejected at the \hbox{$\approx$~22--92\%} level),
we note that the analysis of \xray\ variability for the
quasars studied here is limited by both the short \xray\ exposures
in the rest frame (\hbox{$\approx$~15~min}; see Table~1) and the low
\xray\ fluxes of most of our quasars.

\section{X-ray analysis}

\begin{figure*}[]
\centerline{\includegraphics[angle=0,width=\textwidth]{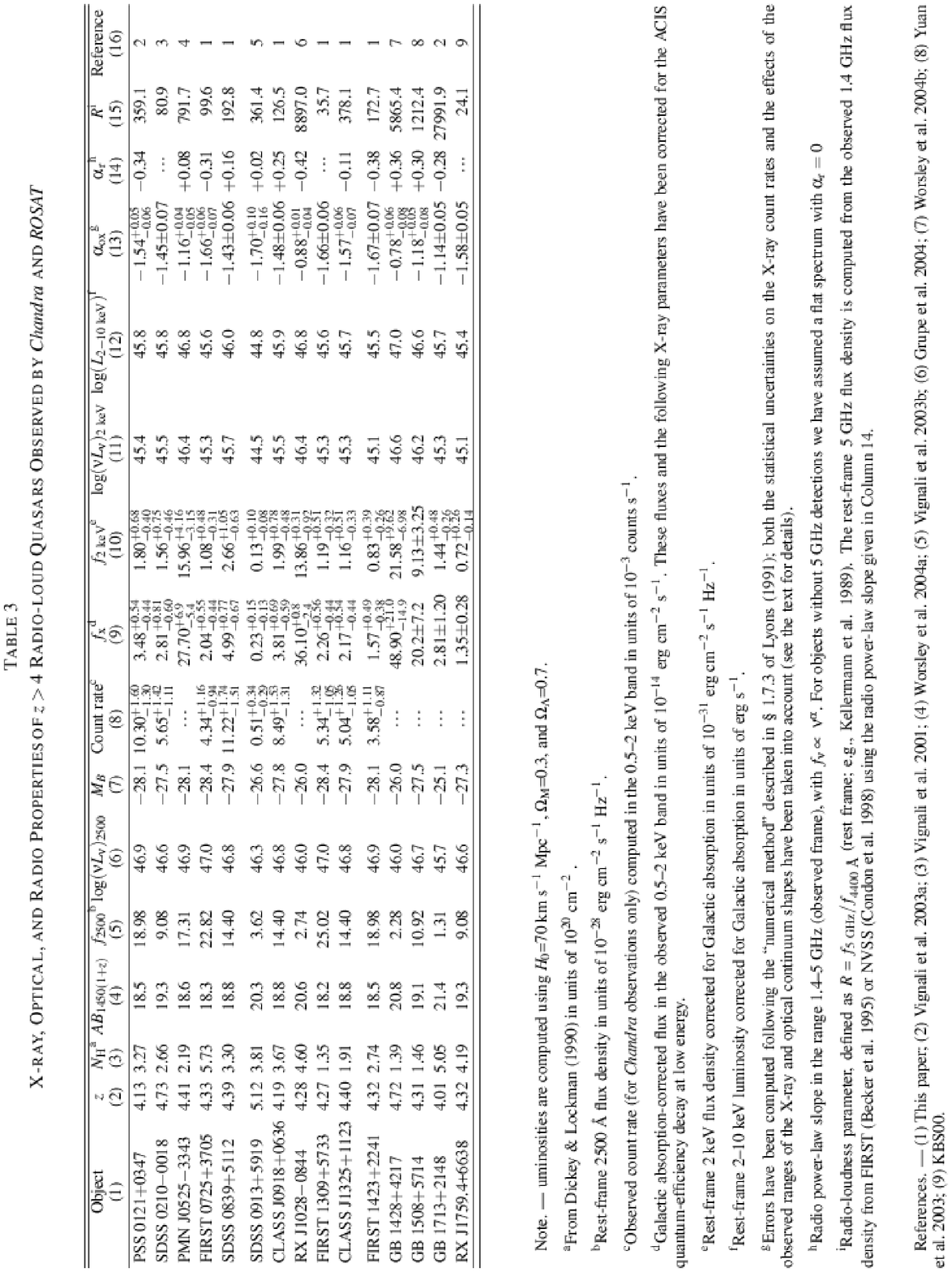}}
\end{figure*}

\begin{figure*}[!t]
\figurenum{4}
\centerline{\includegraphics[angle=0,width=8.5cm]{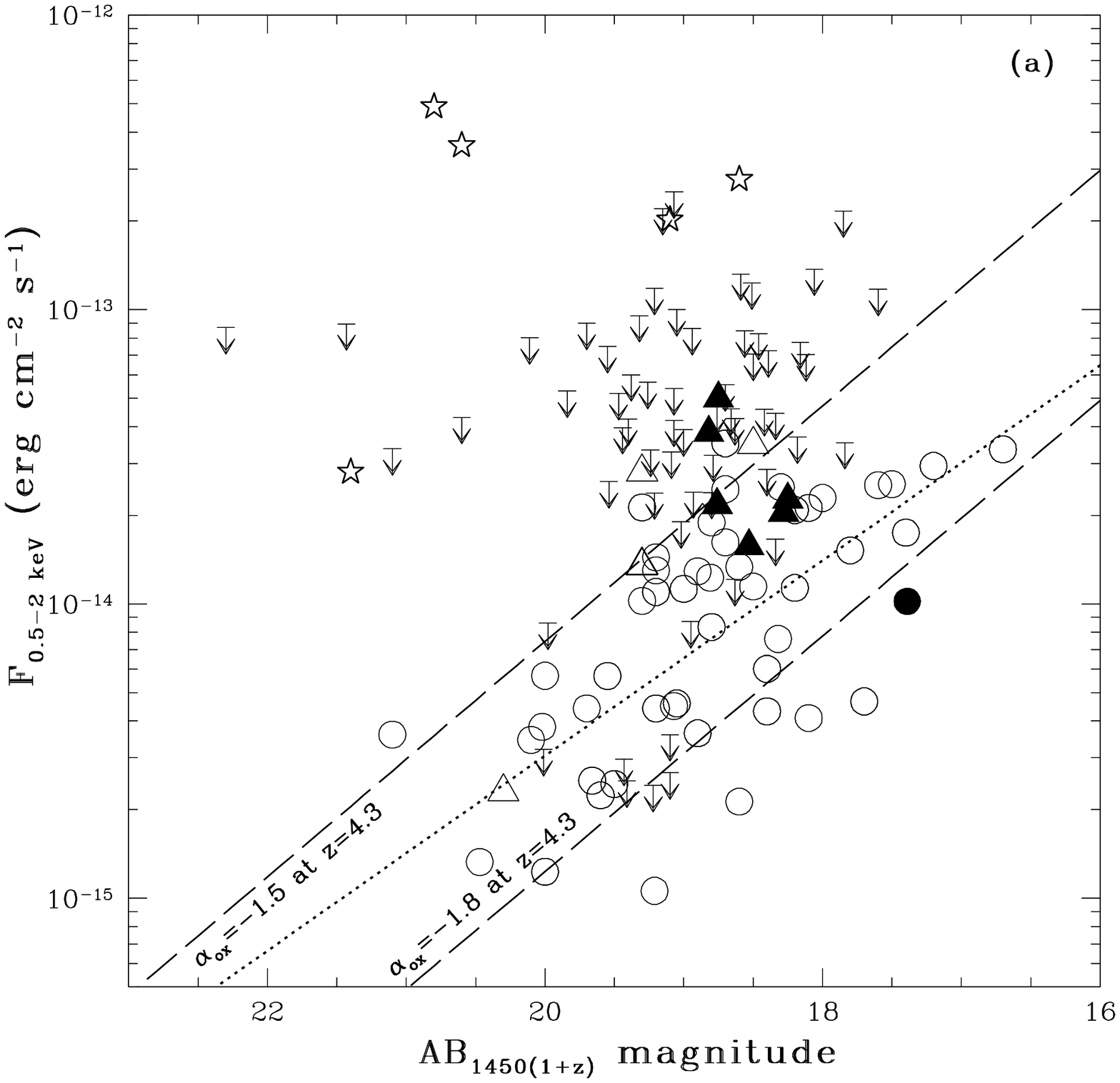}
\includegraphics[angle=0,width=8.5cm]{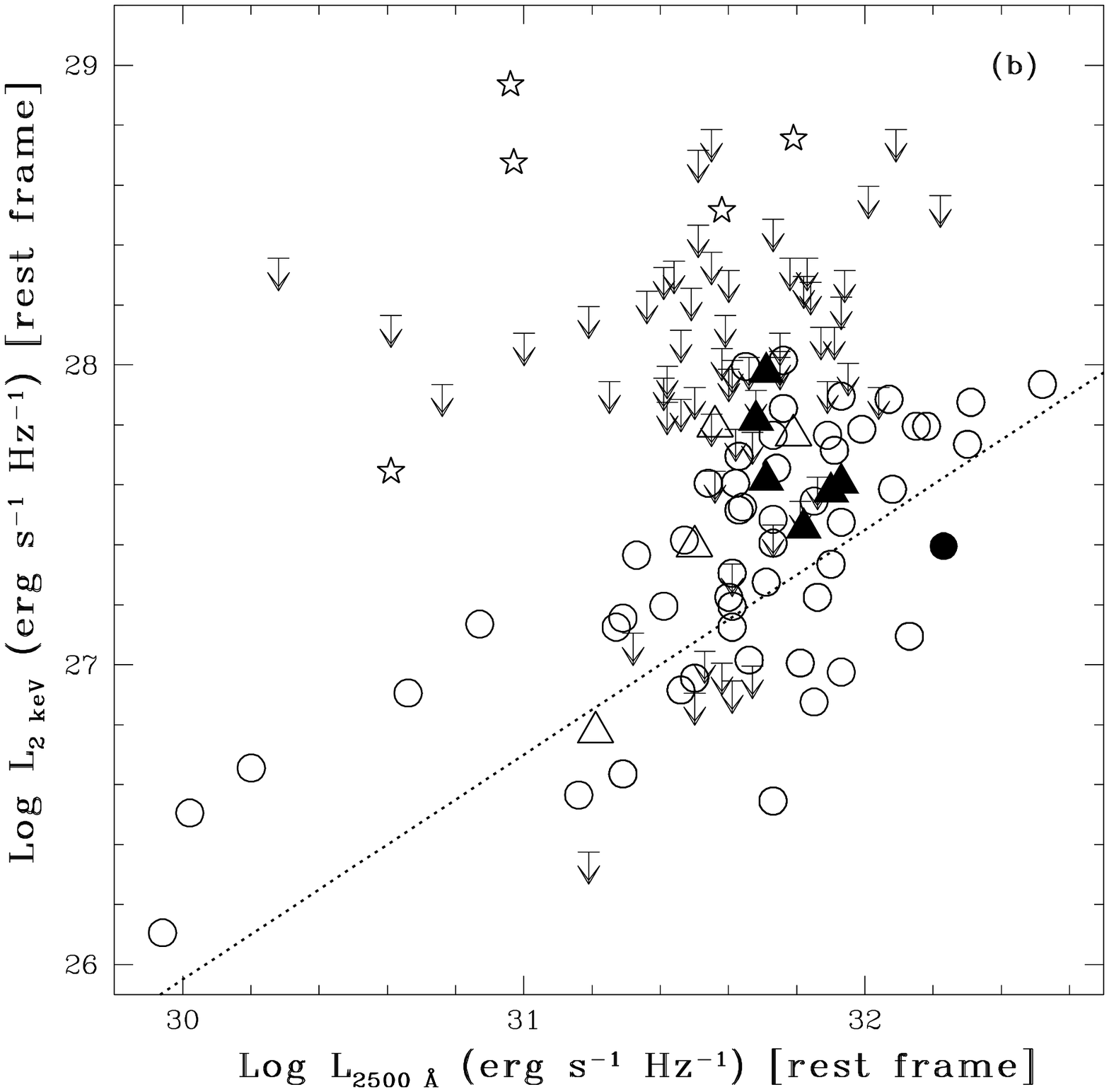}}
\figcaption[bassett.fig4a.ps and bassett.fig4b.ps]{\xray\ -- optical correlations
for $z>4$ RQQs (circles), RLQs (triangles), blazars (stars) and \xray\ upper limits
(downward-pointing arrows, at the \hbox{$\approx 3\sigma$} confidence level).
The new \chandra\ observations presented
in this paper are marked with filled symbols.
(a) Observed-frame, Galactic
absorption-corrected \hbox{0.5--2~keV} flux versus \ab1450\ magnitude for
$z>4$ quasars.  The dashed lines show the
\hbox{$\alpha_{\rm ox} = -1.5$} and \hbox{$\alpha_{\rm ox} = -1.8$} loci at $z =
4.3$ (the average redshift of the present sample).  The dotted
line shows the best-fit correlation reported in $\S$4 [Equation
(2)] of Vignali et al.\ (2003b) for $z \geq
4$ optically selected RQQs.
(b) \xray\ (2~keV) versus UV (2500 \AA) luminosity density
for $z>4$ quasars.  The dotted line indicates the best-fit relationship
for 137 SDSS RQQs in the \hbox{0.16--6.28} redshift range from VBS03.  Clearly,
the RLQs lie significantly above the best-fit line, typically being \xray\
brighter by a factor of $\approx 2$ than RQQs at comparable UV luminosities.
\label{fig4}}
\end{figure*}

The principal \xray, optical, and radio properties of the fifteen
$z>4$ RLQs with \xray\ detections (including those detected prior
to this work) are given in Table~3.
A description is as follows:\\
{\sl Column (1)}. --- The name of the source. \\
{\sl Column (2)}. --- The redshift of the source. \\
{\sl Column (3)}. --- The Galactic column density (from Dickey \& Lockman 1990) in units of \hbox{10$^{20}$ cm$^{-2}$}. \\
{\sl Column (4)}. --- The monochromatic rest-frame \ab1450\
magnitude (defined in $\S$3b of Schneider et al.\ 1989), corrected
for Galactic extinction. When spectroscopically determined values
are not available, \ab1450\ magnitudes have been derived from the
SDSS $i$-magnitudes assuming the empirical relationship
\hbox{$AB_{1450(1+z)}=i - 0.2$}.  For objects without SDSS
detections, we used \hbox{$R$-band} magnitudes from the APM Catalog
(McMahon et al.\ 2002) and the relationship
\hbox{$AB_{1450(1+z)}=R-(0.648)z+3.10$}. We have compiled a list
of all \hbox{$z\ge4$} quasars with similar photometry; this is used in
the analysis throughout this paper.\footnote{This list is available from
http://www.astro.psu.edu/users/niel/papers/papers.html} All of the
above relationships provide reliable $AB_{1450(1+z)}$ estimates
(within \hbox{$\approx$~0.1--0.2} mags)
in the redshift range under consideration.\\
{\sl Columns (5) and (6)}. --- The 2500~\AA\ rest-frame flux
density and luminosity. These were computed from the \ab1450\
magnitude assuming an optical power-law slope of $\alpha=-0.5$
\hbox{($f_{\nu}$ $\propto$ $\nu^{\alpha}$}; Vanden Berk et al.\
2001). The 2500~\AA\ rest-frame flux densities and luminosities
are increased by \hbox{$\approx$~15\%} for an optical power-law slope of
\hbox{$\alpha=-0.79$} (e.g., Fan et al.\ 2001a)
as in Vignali et al.\ (2001, 2003a,b). \\
{\sl Column (7)}. --- The absolute $B$-band magnitude computed
assuming \hbox{$\alpha=-0.5$}. If \hbox{$\alpha=-0.79$} is adopted for
the extrapolation,
the absolute $B$-band magnitudes are brighter by \hbox{$\approx$~0.35} mag. \\
{\sl Columns (8) and (9)}. --- The observed count rate in the
\hbox{0.5--2~keV} band (for \chandra\ observations) and the
corresponding flux ($f_{\rm X}$), corrected for Galactic
absorption. This flux has been calculated using {\sc pimms} and a
power-law model with \hbox{$\Gamma=1.6$}, as derived for samples
of \hbox{$z\approx$~0--2} RLQs (e.g., Reeves \& Turner 2000); see
also $\S$3.2.  Changes of the photon index in the range
\hbox{$\Gamma=1.4$--1.9} lead to only a few percent change in the
measured \xray\ flux.  The soft \xray\ flux derived from the
\hbox{0.5--2~keV} counts is generally similar (to within \hbox{$\approx
5$--15\%}) to that derived using the full-band (0.5--8 keV) counts.

The \xray\ fluxes reported and used in this paper have been
corrected for the ACIS quantum-efficiency degradation at low energy using
the method described in $\S3.2$. The
\hbox{0.5--2~keV} flux correction is \hbox{$\approx$~25\%} for the
\chandra\ observations presented here.\\
{\sl Columns (10) and (11)}. --- The rest-frame 2~keV flux density and luminosity, computed assuming \hbox{$\Gamma=1.6$}. \\
{\sl Column (12)}. ---  The \hbox{2--10~keV} rest-frame luminosity, corrected for Galactic absorption. \\
{\sl Column (13)}. ---  The optical-to-X-ray power-law slope, \aox, defined as
\begin{equation}
\alpha_{\rm ox}=\frac{\log(f_{\rm 2~keV}/f_{2500~\mbox{\rm \scriptsize\AA}})}{\log(\nu_{\rm 2~keV}/\nu_{2500~\mbox{\rm \scriptsize\AA}})}
\end{equation}
where \hbox{$f_{\rm 2~keV}$} and $f_{2500~\mbox{\scriptsize \rm \AA}}$
are the rest-frame flux densities at 2~keV and 2500~\AA,
respectively. The \hbox{$\approx 1\sigma$} errors on \aox\ have been
computed following the ``numerical method'' described in
$\S$~1.7.3 of Lyons (1991). Both the statistical uncertainties on
the \xray\ count rates and the effects of possible changes in the
\xray\ (\hbox{$\Gamma\approx$~1.4--1.9}) and optical
(\hbox{$\alpha\approx$~$-$0.5} to $-$0.9; Schneider et al. 2001)
continuum shapes have been taken into account. Our choice of
\hbox{$\Gamma = 1.6$} and \hbox{$\alpha = -0.5$} accounts for the slight
differences in $f_{2500}$, $f_{2 {\rm keV}}$, and \aox\ values
as compared to those in Vignali et al.\ (2001, 2003a,b). \\
{\sl Column (14)}. --- The radio power-law slope \hbox{($f_{\nu}$
$\propto$ $\nu^{\alpha}$)} between 1.4~GHz and 5~GHz (observed
frame).  The 1.4~GHz flux density is from FIRST or NVSS (Condon et
al.\ 1998), and the 5~GHz flux density is from the GB6 (\hbox{$\delta
> 0$};
Gregory et al. 1996) or PMN (\hbox{$\delta < 0$}; Griffith \& Wright 1993) surveys. \\
{\sl Column (15)}. --- The radio-loudness parameter defined by
Kellermann et al.\ (1989) as \hbox{$R=f_{\rm 5~GHz}/f_{\rm
4400~\mbox{\scriptsize\AA}}$} (rest frame). The rest-frame 5~GHz
flux density was computed from the FIRST or NVSS observed 1.4~GHz
flux density and the radio power-law slope given in column 14.
Where a 5~GHz detection was not achieved, we have assumed a flat
spectrum of \hbox{$\alpha_{\rm r}=0$} to be consistent with the other
\hbox{$z>4$} RLQs. The rest-frame 4400~\AA\ flux density was computed
from the \ab1450\ magnitude assuming an
optical power-law slope of \hbox{$\alpha=-0.5$}. \\
{\sl Column (16)}. --- Recent \xray\ references.

\subsection{\xray\ -- Optical Correlations}

\begin{figure*}[!t]
\figurenum{5}
\centerline{\includegraphics[angle=270,width=\textwidth]{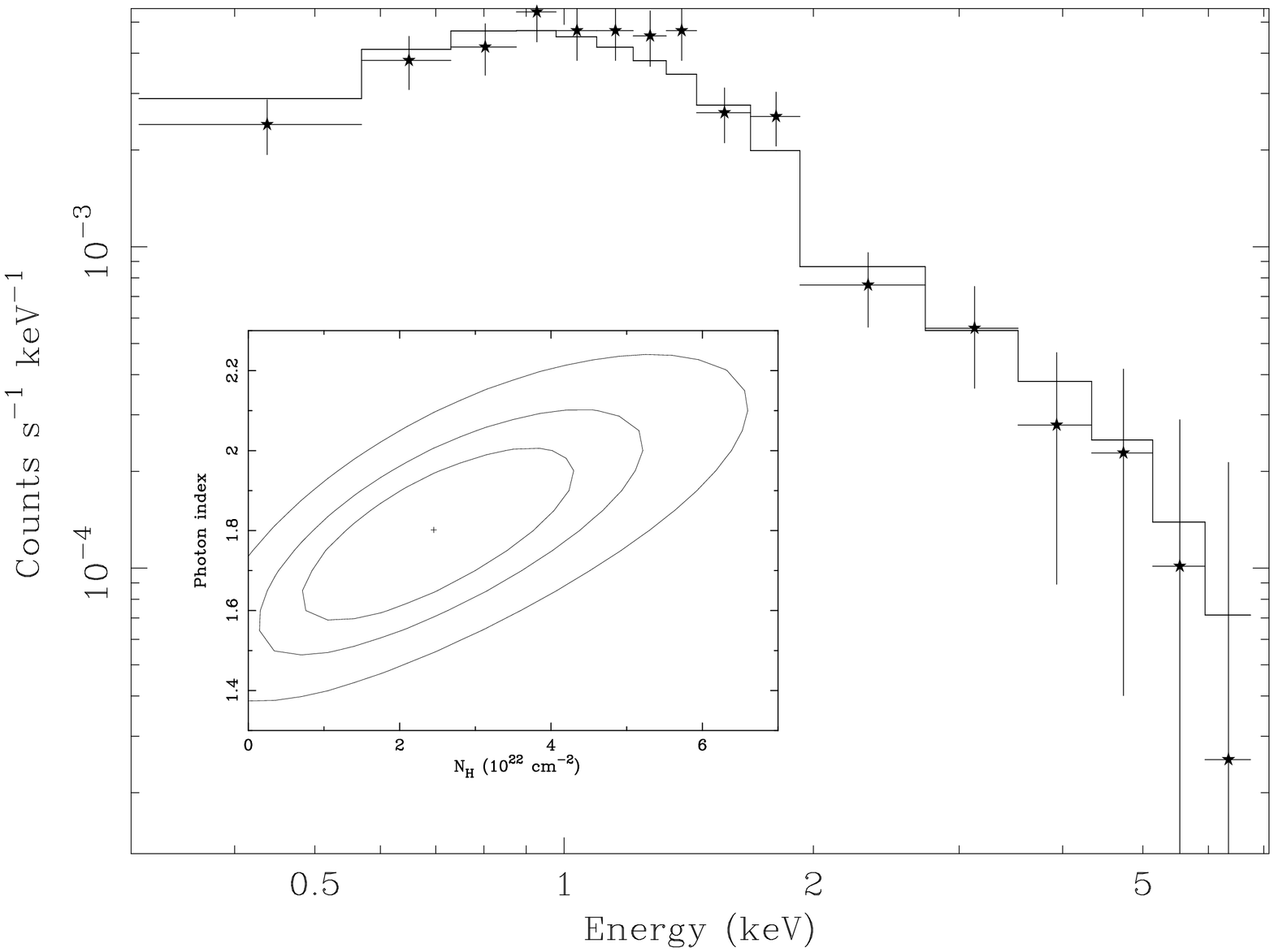}}
\figcaption[bassett.fig5.ps]{Combined spectrum (in the observed frame and used only for presentation
purposes) for the nine \chandra\ observations of $z>4$ RLQs with
moderate-$R$, fitted with a power-law model and
Galactic absorption (see $\S$3.2 for details).  The 68, 90, and
99\% confidence regions for the photon index and intrinsic column
density obtained using the Cash statistic are shown in the inset.
\label{spec}}
\end{figure*}

With nearly one hundred published \xray\ detections of quasars at
$z>4$, the typical \xray\ fluxes of \hbox{$z \approx$~4--5} quasars are
now well defined. Plots of \xray\ flux versus $AB_{1450(1+z)}$
magnitude are often constructed for distant quasars and are
useful for planning future observations, as they
reflect directly observable quantities.  Figure~4a is such a plot,
depicting the observed-frame \hbox{0.5--2~keV} flux versus \ab1450\ for
quasars at $z>4$ with \xray\ detections and upper limits. A strong
correlation exists between these quantities, and the best-fit
relationship derived in Vignali et al.\ (2003b) for $z>4$ RQQs is
shown as a dotted line.  It is immediately apparent that the RLQs in this
sample (depicted as filled triangles) are generally brighter
in X-rays than RQQs (circles) with comparable optical magnitudes, all lying
above the best-fit relationship.  The RQQ FIRST~0747$+$2739
(filled circle) lies below the best-fit.

Extending much earlier work, VBS03 found that
the \xray\ and broad-band properties of RQQs do not significantly evolve
with redshift.  VBS03 found that, independent of redshift in the range
\hbox{$z \approx 0.2$--6.3}, the \xray\ luminosities (\Lx) of RQQs are highly
correlated (9.3$\sigma$ in VBS03) with their ultraviolet (UV) luminosities (\Luv).
These quantities are plotted for the \xray\ detected $z>4$ quasars
in Figure~4b, along with the best-fit relationship for
RQQs from VBS03 (adapted for an optical slope of \hbox{$\alpha = -0.5$}).
Clearly, the RLQs (triangles) lie significantly above the best-fit
line for RQQs (circles), with systematically higher \xray\
luminosities by a factor of $\approx 2$.  These are the same trends exhibited by
samples of RLQs at low-to-moderate redshift, in which the
additional \xray\ emission is attributed to a jet-linked
component.  In general, the excess \xray\ emission tends to
correlate with radio-loudness (see, e.g., Figure~6), lending additional
support to this interpretation.   We further investigate the implications of these
trends and the nature of the jet-linked \xray\ emission in $\S$3.3

\subsection{Joint Spectral Fitting}

To provide tighter constraints on the average \xray\ spectral
properties of $z>4$ RLQs, we have performed a joint spectral
analysis of the nine \hbox{moderate-$R$} (\hbox{$R \approx 40$--400}) RLQs
observed by \chandra; these have an average redshift of \hbox{$\langle z \rangle
= 4.43$}. Three of these objects have been previously published:
PSS~0121$+$0347, Vignali et al.\ (2003a); SDSS~0210$-$0018,
Vignali et al.\ (2001); and SDSS~0913$+$5919, Vignali et al.\
(2003b). Source counts have been extracted from 2\arcsec\ radius
circular apertures centered on the \xray\ position of each quasar.
The background has been taken from annuli centered on the sources,
avoiding the presence of nearby faint \xray\ sources.
Only one object is characterized by low
counting statistics (SDSS~0913$+$5919, which has 6 counts); all of
the others have \hbox{$\approx 20$--80} counts in the \hbox{0.5--8~keV} band.
The sample does not appear to be biased by the presence of a few
objects with atypically high signal-to-noise ratios (see Table~2),
and the removal of any one of these objects does not produce
significantly different results.

We have corrected each spectrum for the quantum-efficiency degradation of
ACIS at low energies using the correction supplied via
the \chandra\ Calibration Database ({\sc caldb}), Version 2.26.  While
the degradation is not severe above $\approx 0.7$~keV, it could
significantly affect estimates of \xray\ absorption if not
properly corrected.  We have also repeated our analysis using the
{\sc acisabs} correction,\footnote{
See http://www.astro.psu.edu/users/chartas/xcontdir/xcont.html}
based on data from the external calibration source aboard \chandra\
(the {\sc caldb} correction is derived from gratings data).  Due to
differences in the extrapolation of the time-dependent quantum efficiency between the
models (see Marshall et al.\ 2004), the {\sc acisabs} correction
leads to fits with a slightly lower absorption column and photon
index, but the results are generally consistent.

The joint spectral fitting of the nine unbinned quasar
spectra was carried out using the Cash statistic (Cash 1979),
which is well suited to low-count sources (e.g.,
Nousek \& Shue 1989).  Since the Cash statistic is applied to the
unbinned data, it allows us to retain all available spectral
information.  We can also assign to each source its own
Galactic absorption and redshift. The resulting joint spectrum (in
the \hbox{$\approx 2$--40~keV} rest-frame band), while derived from only
$\approx 350$~counts, is reasonably good due to the extremely low
background of \chandra\ (there are only $\approx 5$~background
counts in all source cells). Errors in the following analysis
are quoted at 90\% confidence for one interesting parameter.

Fitting a power-law model with
Galactic absorption using {\sc xspec}, we find a good fit (shown in
Figure~\ref{spec}) with an average photon index of \hbox{$\Gamma
 = 1.65\pm0.15$}.
The fitted photon index is
consistent with that derived using the
band ratios for the six RLQs in Table~2, from which we obtain a
weighted mean of \hbox{$\Gamma  = 1.77 \pm 0.16$}. It is
also consistent with previous studies of \hbox{$z \approx 0$--4} RLQs
(e.g., Brinkmann et al.\ 1997;
Reeves \& Turner 2000) which have found that core-dominated, flat-spectrum RLQs
typically have \hbox{$\Gamma \approx$1.4--1.9}.

To search for evidence of intrinsic \xray\ absorption in the
above sample of nine moderate-$R$ RLQs, we added to the joint spectral
model a redshifted neutral absorption component (the {\sc xspec}
model {\it zphabs}) with the redshift set to that of each
source.  With this model, we find tentative evidence ($\simgt$90\% confidence)
for absorption in excess of the Galactic $N_{H}$, with a best-fit value
of \hbox{$N_{H} = 2.4^{+2.0}_{-1.8} \times 10^{22}$~cm$^{-2}$} (errors at 90\%
confidence for one interesting parameter).  This fit produces
a somewhat higher photon index, \hbox{$\Gamma = 1.80^{+0.23}_{-0.25}$},
which is still consistent with the typical range for RLQs from above.
The confidence contours in
the \hbox{$N_{H}$--$\Gamma$} plane are shown in
Figure~\ref{spec}.

From a large sample of \asca\ observations, Reeves \& Turner (2000) confirmed
earlier suggestions by Cappi et al.\ (1997) and Fiore et al.\ (1998)
of a significant \hbox{$N_{H}$--$z$} correlation for RLQs, finding
that \hbox{$z \approx 2$--4} RLQs have typical column densities
of \hbox{a few $\times 10^{22}$~cm$^{-2}$} (see their Figure 9).
Also, for two of the $z>4$ blazars,
strong intrinsic \xray\ absorption (\hbox{$10^{22}$--10$^{23}$~cm$^{-2}$})
has been suggested by spectral flattening at low energies:
PMN~0525$-$3344 (Worsley et al.\ 2004a) and GB~1428$+$4217
(e.g., Fabian et al.\ 2001; Worsley et al.\ 2004b).  Our
findings are generally consistent with these claims and extend this
result to the majority of RLQs with moderate radio-loudness (the
average radio-loudness of the objects in the above
samples is \hbox{$\langle R \rangle \approx 10\, 000$}, while that for
the present sample is \hbox{$\langle R \rangle \approx 200$}).

\subsection{The Nature of the Jet-Linked Component}
The enhanced radio and \xray\ emission of RLQs in comparison to
RQQs is usually attributed to a relativistic jet-linked component.
For core-dominated, flat-spectrum RLQs, the majority of the \xray\
emission is presumably from the small-scale jet close to the AGN,
which, aided by Doppler boosting, accounts for the increased \xray\ luminosities
and flatter \xray\ spectra of RLQs as a whole (e.g., Worrall et al.\ 1987;
Wilkes \& Elvis 1987).
For more extended (kpc scale) \xray\ jets, there
has been considerable debate over the process responsible for
the \xray\ emission.  In several cases, low optical luminosities, which
preclude pure synchrotron emission from a single electron population,
have led to synchrotron models with more complicated electron energy
distributions (e.g., Stawarz et al.\ 2004) or models based on some
form of IC scattering.  In the case of IC emission, individual
observations of \xray\ jets lead to differing
conclusions as to the primary source of IC seed photons (see Harris \&
Krawczynski 2002 for a review). The
obvious choices are the synchrotron photons themselves
(synchrotron self-Compton -- SSC), UV photons from the accretion
disk, or cosmic microwave background photons (IC/CMB); the relative
importance of these sources may depend on factors such as the quasar's
environment, orientation, and the bulk relativistic motion of
the jet.  Both SSC and IC from UV photons produced in the AGN
presumably dominate close to the core of the quasar, while IC/CMB
may dominate at large distances.  At high redshift IC/CMB is
particularly relevant since, as pointed out originally by Rees \& Setti
(1968) and recently by Schwartz (2002), the cosmological decrease
in surface brightness as $(1+z)^{-4}$ is compensated for by a
$(1+z)^{4}$ increase in the energy density of the CMB.
Above \hbox{$z \approx 3$--4}, \xray\ jets dominated by the IC/CMB process
could even outshine their parent quasars (see Figure~2 of Schwartz 2002).

\figurenum{6}
\centerline{\includegraphics[angle=0,width=8.5cm]{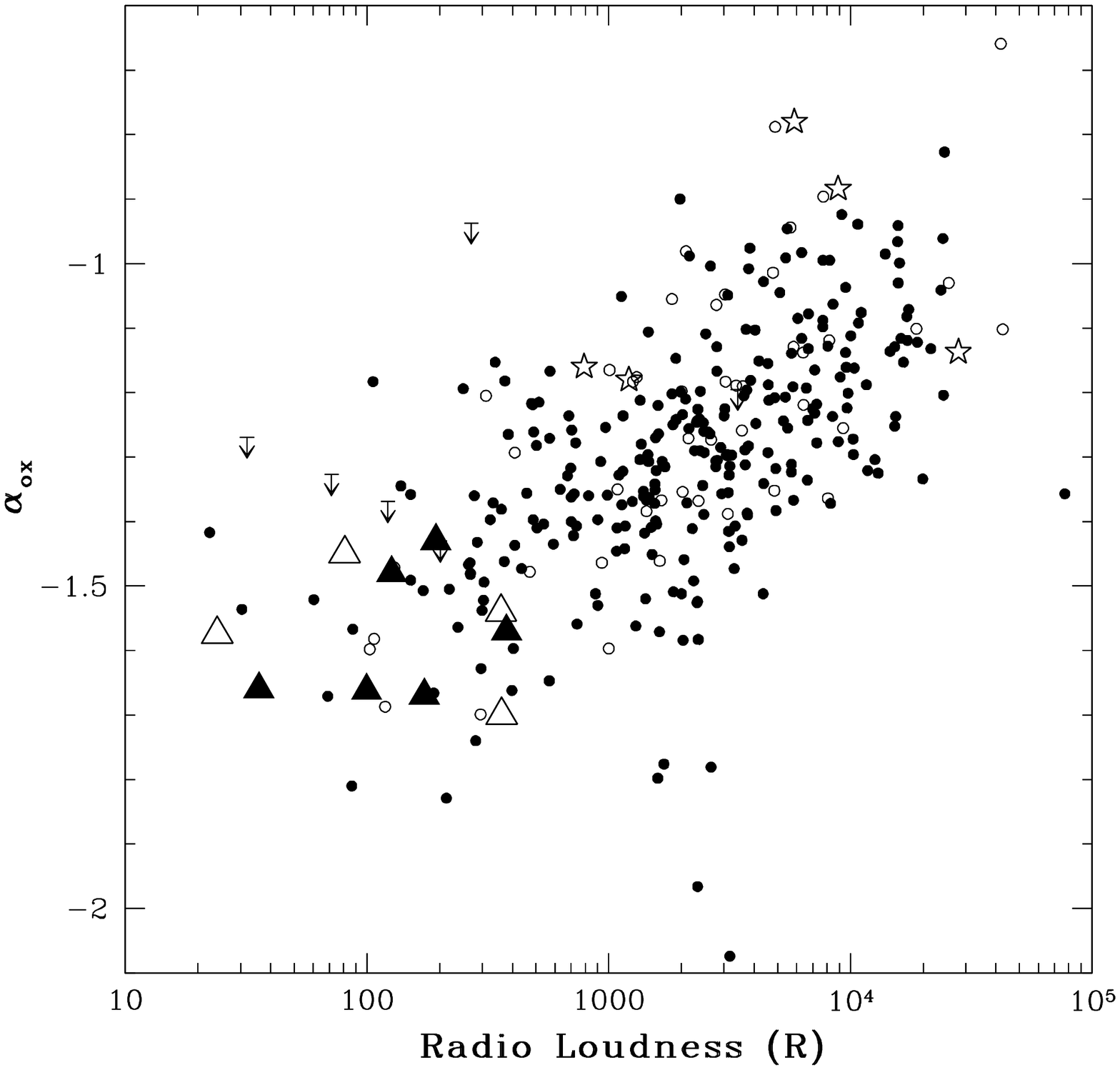}}
\figcaption[bassett.fig6.ps]{Broad-band spectral index \aox\ versus
radio-loudness ($R$) for samples of flat-spectrum RLQs at differing redshifts.  The
small circles represent the RLQs from the radio-selected sample of
Brinkmann et al.\ (1997), at redshifts of \hbox{$z \approx 0.2$--2} (filled
circles) and \hbox{$z \approx 2$--4} (open circles).  The large symbols
represent $z>4$ RLQs with \xray\ detections and upper limits
(downward pointing arrows).  We see no significant evidence for \xray\
enhancement of RLQs at high redshift, as both the $z>4$ \hbox{moderate-$R$} RLQs
(triangles) and the blazars (stars) fall well within the distribution
of objects at lower redshift.  The new \chandra\ observations from
this paper are plotted as filled triangles. \label{brink}} \centerline{} \centerline{}

As established in $\S$3.1, our $z>4$ RLQs are $\approx 2$ times
\xray\ brighter than RQQs with comparable optical luminosity, so
the X-rays we have detected appear to be largely from a jet-linked
component.  We do not, however, observe any evidence for spatially
extended emission in the \chandra\ images (see $\S$2.3), nor do
the FIRST radio images show any obvious spatial structure (note
that the sub-arcsecond angular resolution of \chandra\ is better than the
$\approx 5$\arcsec\ resolution of the FIRST survey). In Figure~\ref{brink} we
have plotted the broad-band spectral index $\alpha_{\rm ox}$
versus radio loudness for our observations of $z>4$ RLQs
(triangles) along with the Brinkmann et al.\ (1997) sample of
radio-selected flat-spectrum RLQs which spans the $z<2$ (solid
circles) and $z = 2$--4 (open circles) redshift ranges. Although
the number of moderately radio-loud RLQs in the Brinkmann et al.\
(1997) sample is limited, it is clear that the $z>4$ RLQs are
comparable in their \xray\ and broad-band properties to those at
low-to-moderate redshift.  The $z>4$ blazars (star symbols) also
lie within the observed range of these parameters for comparable
low-redshift objects.  We have performed a similar comparison of
our $z>4$ RLQs to the low-to-moderate redshift sample of
flat-spectrum RLQs presented in Fig.~4 of Fabian et al.\ (1999)
and reach the same conclusion.

Figure~6 has implications for jet \xray\ emission mechanisms in RLQs.
In particular, the
observations provide evidence that the IC/CMB process does not
dominate the total \xray\ production of these objects; if it did,
then the the arguments of Schwartz (2002) predict that these jets
at $z>4$ should be
as bright or brighter in X-rays than the quasar cores.  Such an enhancement
would lead to a systematic flattening of \aox\ by \hbox{$\approx$~0.1--0.2} relative to
comparable objects at low redshift, which we do not observe.  Rather, it
seems that the X-rays are probably made primarily on small scales where the
quasar photon field dominates the CMB photon field.  This is consistent
with expectations for low-redshift flat-spectrum RLQs, where on small scales
SSC and external IC scattering from AGN seed photons are thought
to dominate.

Still, the question remains:  if we believe that much of the \xray\
emission from these objects is associated with a jet-linked
component on small scales, then why do we not see the extended
IC/CMB jet, which should be significantly enhanced by the
increased energy density of the CMB at high redshift?
Schwartz (2002) uses the jets from the quasars 3C~273 ($z=0.158$)
and PKS~0637$-$752 ($z=0.654$)
as examples of IC/CMB jets (at projected distances of roughly 25~kpc
and 75~kpc, respectively) that should outshine their parent quasars
above $z\approx 4$.  Both of these quasars have flat (core) radio
spectra similar to those of our objects, and the predicted \chandra\ count
rates of the cores as they would appear at $z\approx 4$ (see Figure~2
of Schwartz 2002) of a few$\times 10^{-3}$~cts~s$^{-1}$
are consistent with our observations.  However, 3C~273 and PKS~0637$-$752 are
more radio loud than our quasars by nearly an order of magnitude,
with $R\approx 1500$ and 2200, and so it may be that
differences in either orientation or intrinsic jet power may
account for our observations.\footnote{
The lower radio loudnesses of our objects may indicate that they
have jets directed along a larger angle to the line of sight, thus making
it less likely to observe beamed IC/CMB X-rays, which require a fast jet close to the line
of sight (e.g., Harris \& Krawczynski 2002).  However, the lower
radio loudnesses may also
result from differences in intrinsic jet power, and these
effects are not easy to separate.
}
But even in the case of the
recently discovered \xray\ jet from the $z=4.3$ blazar
GB~1508$+$5714, where the IC/CMB process has been invoked
(Siemiginowska et al.\ 2003; Yuan et al.\ 2003; Cheung 2004),
we do not see such a dramatic enhancement. This
jet only accounts for $\approx3$\% of the quasar luminosity and
was discovered in a \chandra\ observation with a lengthy (89~ks)
exposure time.  To test our sensitivity to detecting such jets,
we randomly selected 800~s segments of the
GB~1508$+$5714 data, chosen to have $\approx40$~ counts in the
full \hbox{0.5--8~keV} band, matching the average number of counts
in our \hbox{$\approx$5~ks} \chandra\ snapshot observations.  We analyzed these
images following the procedure of $\S$2.3 and did not find clear
evidence for extent in any case.  Inspecting the raw and
adaptively smoothed images by eye, only in
\hbox{$\approx$10--20\%} of the fields could we observe hints of
the jet.  This is not surprising, as at $\approx 3$\% of
the quasar luminosity we expect only 1--2 counts from the jet in
each field.  So if our moderate-$R$ RLQs have jets similar in
nature to that of GB~1508$+$5714 (perhaps at a larger viewing
angle to account for the difference in radio-loudness) then we
would not expect to detect extended jets from these snapshot
images.

Overall, from this preliminary sample, it appears
that the small-scale jet components of flat-spectrum RLQs do not change
significantly between redshifts 0 and 5.  This result is broadly in accordance with
recent studies (e.g., VBS03) which imply
that the disk/corona components of RQQs do not significantly evolve with
redshift in the range \hbox{$z \approx 0$--5}.  Apparently, RLQ jets are also
almost completely built-up systems at \hbox{$z \approx$4--5}.

\subsection{FIRST~0747$+$2739}
Although the $z=4.11$ quasar FIRST~0747$+$2739 is radio-quiet (a faint
detection in the FIRST survey gives a radio-loudness of $R\approx 2$),
it is one of the most optically luminous quasars at $z>4$ with \hbox{$M_{B}=-29.2$}
(compare with Figure~1 of Vignali et al.\ 2003a).
Richards et al.\ (2002) found an overabundance of C {\sc iv} absorption systems
in the spectrum of FIRST~0747$+$2739, with at least 14 independent absorption
systems spanning a redshift range of \hbox{$\Delta \, z \approx 1$} longward of the Ly$\alpha$
forest.  They concluded that the absorption is most likely intrinsic to the
quasar, resulting from multiple high-velocity absorption systems at the spatial,
density, or temporal edge of a BAL region.
FIRST~0747$+$2739 may be a quasar in the transition region
between BAL and normal quasars.

From the \chandra\ observation of FIRST~0747$+$2739, we measure an observed-frame
\hbox{0.5--2~keV} flux of \hbox{$1.02\times 10^{-14}$~erg~cm$^{-2}$~s$^{-1}$}, from which
we derive a rest-frame \hbox{2--10~keV} luminosity of \hbox{$2.3\times 10^{45}$~erg~s$^{-1}$} and
an optical-to-\xray\ spectral index of \hbox{\aox$=-1.86$}. Based on a significant
correlation between C {\sc iv} absorption and \xray-weakness,
Brandt, Laor, \& Wills (2000) have suggested that \xray\ absorption associated with a BAL-type
outflow is the primary cause of \xray-weakness in $z<0.5$ blue AGN, and recent
results suggest that this is likely the case for $z>4$ quasars as well (see Vignali et al.\ 2003b
and references therein).  Despite its significant C {\sc iv} absorption,
FIRST~0747$+$2739 does not
appear to be notably X-ray weak when compared to other highly
luminous quasars (see, e.g., Figure~3 of Brandt, Schneider, \& Vignali 2004).
However, since we are sampling highly penetrating X-rays with
rest-frame energies of \hbox{$\approx$~3--40~keV} (and have a limited
number of photons), the X-ray column density constraints are not
tight. Spectral fitting as described in \S3.2 indicates that
intrinsic neutral absorption with \hbox{$N_{H} < 5.5 \times 10^{22}$~cm$^{-2}$}
is consistent with our data.
Of course, even larger column densities can be accommodated for
ionized X-ray absorption (as is likely expected from the observation
of C {\sc iv} absorption). Thus, our results are at least consistent
with expectations from Brandt et al.\ (2000), even if
they cannot directly support these expectations.

\section{Summary and Future Work}
We have presented new \chandra\ detections of six RLQs and one RQQ
at \hbox{$z\approx$~4.1--4.4}, thereby significantly enlarging the number of
$z>4$ RLQs with X-ray detections. Our RLQ targets have moderate $R$
parameters of \hbox{$\approx$~40--400}; they are more representative of the
overall RLQ population than many of the \hbox{high-$R$} ($R>500$) RLQs at
$z>4$ previously studied in X-rays. Our main results are the following:
\begin{enumerate}
   \item Moderate-$R$ RLQs at $z>4$ have enhanced X-ray emission relative
     to RQQs of the same redshift and optical luminosity.
     The degree of enhancement, a factor
     of $\approx 2$ in X-ray luminosity, is consistent with that
     observed at low redshift for comparable objects and is presumably
     associated with a jet-linked emission component. As in the local
     universe, the X-ray-to-optical flux ratios of \hbox{moderate-$R$} RLQs at $z>4$ are
     intermediate between those of RQQs and those of \hbox{high-$R$} RLQs.
   \item Joint X-ray spectral analyses of moderate-$R$ RLQs show that,
     on average, their spectra can be fitted with a power-law model
     and Galactic absorption; the average best-fit power-law photon
     index of \hbox{$\Gamma = 1.65\pm0.15$} is
     consistent with that for comparable objects at low redshift. Adding
     an absorption component to the fit, we
     find tentative evidence (90\% significance) for excess
     intrinsic X-ray absorption on average, and our
     best-fit estimate of the cold absorption column of
     \hbox{$N_{H}=2.4^{+2.0}_{-1.8} \times 10^{22}$~cm$^{-2}$} is consistent
     with previous claims of increased absorption toward
     high-redshift RLQs.  The somewhat higher
     best-fit photon index of \hbox{$\Gamma = 1.80^{+0.23}_{-0.25}$} in this model
     is still consistent with previous estimates for RLQs at low redshift.
   \item We do not detect extended X-ray emission associated with jets
     from any of our RLQs, which, if dominated by the IC/CMB process as
     suggested, should be significantly enhanced at $z>4$.  If the jets in these
     objects are similar to the more modest \xray\ jet recently discovered
     from the $z=4.3$ blazar GB~1508$+$5714 (which accounts for
     only $\approx 3$\% of the quasar \xray\ luminosity), we expect that the jets
     are below the sensitivity limit of our snapshot observations.
   \item Overall, our data are consistent with the idea that the dominant
     X-ray production mechanisms of $z>4$ and $z\approx 0$ RLQs are
     similar.
   \item  Richards et al.\ (2002) suggested that an abundance of C {\sc iv} absorption
     systems in the optical spectrum of FIRST~0747$+$2739 are intrinsic to the quasar
     and that it represents a class
     of objects in a transition phase between normal and BAL quasars.
     From our observations, FIRST~0747$+$2739 does not appear to be
     notably \xray\ weak, but we cannot rule out neutral intrinsic
     absorption columns with \hbox{$N_{H} < 5.5 \times 10^{22}$~cm$^{-2}$}.
\end{enumerate}
Finally, several of the RLQs we have detected with \chandra\ are among
the X-ray brightest non-blazars known at $z>4$ (see Figure~4a). These
are prime targets for follow-up X-ray spectroscopy with \xmm\ (with
the caution that in some cases nearby bright sources may contaminate the results). Such
spectroscopy can reveal if the majority population of RLQs at
high redshift shows intrinsic X-ray absorption.

\acknowledgements
We thank A.\ Celotti, M.\ Cirasuolo, Z.\ Ivezi\'{c}, D. Schwartz,
B.\ Wilkes, and D.\ Worrall for helpful discussions or sharing data, and we thank the
referee for a constructive report.
Support from NASA grant NAS8-01128 (GPG, Principal Investigator),
NSF CAREER award AST-9983783 (LCB, WNB, CV), the Penn State
President's Fund for Research (LCB),  NASA LTSA grant
NAG5-13035 (WNB, DPS), and NSF grant 03-007582 (DPS)
is gratefully acknowledged.  CV also acknowledges support
from the Italian Space Agency under contract ASI~I/R/057/02.

\end{document}